\documentclass{emulateapj}
\usepackage{amsmath}
\usepackage{graphicx}
\usepackage{amsfonts}
\usepackage{natbib}
\usepackage{color}
\usepackage{epstopdf}
\usepackage{graphicx}
\usepackage{multirow}
\usepackage{longtable}
\usepackage{rotating}

\begin{document}

\title[$\gamma$-ray QPM in $\gamma$-ray AGNs]{Searching for Quasi-Periodic Modulations in $\gamma$-ray Active Galactic Nuclei}
\author{Peng-fei~Zhang\altaffilmark{1,2}, Da-hai~Yan\altaffilmark{3,4,}, Jia-neng~Zhou\altaffilmark{5}, Jian-cheng~Wang\altaffilmark{3,4,}, Li~Zhang\altaffilmark{1}}
\altaffiltext{1}{Department of Astronomy, School of Physics and Astronomy, Key Laboratory of Astroparticle Physics of Yunnan Province, Yunnan University, Kunming 650091, People's Republic of China; zhangpengfei@ynu.edu.cn}
\altaffiltext{2}{Key Laboratory of Dark Matter and Space Astronomy, Purple Mountain Observatory, Chinese Academy of Sciences, Nanjing 210008, People's Republic of China}
\altaffiltext{3}{Key Laboratory for the Structure and Evolution of Celestial Objects, Yunnan Observatory, Chinese Academy of Sciences, Kunming 650011, People's Republic of China; yandahai@ynao.ac.cn}
\altaffiltext{4}{Center for Astronomical Mega-Science, Chinese Academy of Sciences, 20A Datun Road, Chaoyang District, Beijing, 100012, People's Republic of China}
\altaffiltext{5}{Shanghai Astronomical Observatory, Chinese Academy of Sciences, 80 Nandan Road, Shanghai 200030, People's Republic of China; zjn@shao.ac.cn}

\begin{abstract}
We perform a systematic search of quasi-periodic variabilities in $\gamma$-ray active galactic nuclei (AGNs) in the third
\emph{Fermi} Large Area Telescope source catalog (3FGL).
We employe two techniques, Lomb-Scargle Periodogram (LSP) and Weighted Wavelet Z-transform (WWZ),
to obtain power spectra of $\gamma$-ray light curves covering from 2008 August to 2016 December.
The results show that besides several objects that have been reported in previous works,
an additional source, the FSRQ PKS 0601-70 has a possible quasi-periodic variability of 450 days
in its $\gamma$-ray light curves with the significance of $>3\sigma$.
The physical implications of our findings are discussed.
\end{abstract}

\bigskip
\bigskip


\section{INTRODUCTION}
\label{sec:intro}

The Large Area Telescope (LAT) onboard the \emph{Fermi} satellite is sensitive to $\gamma$-rays from $\sim$20 MeV to 500 GeV \citep{Atwood2009}.
Before the launch of the \emph{Fermi}  satellite two classes of AGNs were known to emit $\gamma$-ray photons: blazars and radio galaxies, both possessing relativistic jets. Their host galaxies are giant elliptical galaxies.  \emph{Fermi}-LAT discovered the third class of  $\gamma$-ray AGNs: radio-loud narrow-line Seyfert 1 galaxies \citep[NLSy1s;][]{Abdo2009,Liao2015}.

Blazars, with their relativistic jets pointing to the Earth \citep{Urry2011}, are the most numerous extragalactic $\gamma$-ray sources in \emph{Fermi}-LAT sources \citep{Acero2015}.
Radiations from blazars cover from radio frequency, optical band, to $\gamma$-ray energies \citep{Maraschi1992,Tavecchio1998,Finke2008,Yan2014},
and are variable in entire electromagnetic bands on a variety of timescales and
amplitudes. According to the features of their optical emission lines, blazars are usually divided into two classes: BL Lac objects (BL Lacs) with weak or even no emission lines and flat spectrum radio quasars (FSRQs) with strong emission lines. 

Radio galaxies are these AGNs with their jets pointing away from our line of sight.
A few radio galaxies were detected by \emph{Fermi}-LAT \citep{Acero2015}. They are lower-luminous than blazars.
\emph{Fermi}-LAT revealed the Mpc jet structure of the radio galaxy Centaurus A \citep{Abdo2010}.

Radio-loud NLSy1s are not typical seyfert galaxies because they possess relativistic jets, which are a relatively rare class of AGNs.
$\gamma$-ray NLSy1s may be some FSRQ-like objects but with lower black hole masses (as well as jet powers) and higher accretion rates \citep[e.g.,][]{Da,Fo,Sun}.
Differing from the host galaxies of blazars and radio galaxies, the host galaxies of radio-loud NLSy1s are spiral galaxies.

\emph{Fermi}-LAT has been collecting $\gamma$-rays more than 8 years, which enables us to construct long-term $\gamma$-ray light curves.
Long-term $\gamma$-ray light curves provide abundant information on jet physics and central super-massive black holes (SMBH).
Here we focus on searching for quasi-periodic variabilities in $\gamma$-ray AGNs.

Quasi-periodic oscillations (QPOs) are important tools  for studies of BH-jet systems.
X-ray QPOs are well observed in X-ray binaries, and are widely believed to be related to the accretion
in the innermost stable circular orbit around BHs \citep{Remillard2006}.
Up to now,  QPOs have been tried to search for in multiwavelength light curves of AGNs.
Many authors claimed the findings of quasi-periodic signals in  AGNs with a wide range of periods from minutes to years \citep{Bai1998,Bai1999,Rani2010,Zhang2012,Lin2013,Zhang2014,Bhatta2016,Li2017,Hong2017,zhangjin}.
Among them, the most compelling case is the BL Lac OJ 287, which has been monitored in optical band more than a century,
and a periodic signal with the period of $\sim$ 12 year was found in its optical light curve \citep{Kidger1992,Valtonen2006}.
In X-ray band, possible QPOs have been reported in NLSy1s (e.g., RE J1034+396, 1H 0707-495 and Mrk 766)
by \citet{Gierlinski2008,Pan2016,Zhang2017d,Zhang2018}, and the periods of X-ray QPOs are usually several hours.
Very recently, $\gamma$-ray QPOs in blazars have been studied by \citet{1553,Sandrinelli2014,Sandrinelli2016a,Sandrinelli2016b,Sandrinelli2017,Zhang2017a,Zhang2017b,Zhang2017c}.
These authors claimed possible $\gamma$-ray QPOs in several blazars
(e.g., PG 1553+113, PKS 2155$-$304, PKS 0537$-$441, PKS 0301$-$243, and PKS 0426$-$380) with the periods of $\sim$1 to 3 years.

In this work, a systematic searching for $\gamma$-ray QPOs in  \emph{Fermi}-LAT AGNs is performed.
New $\gamma$-ray QPO candidates are found.

\section{Results}
\label{results}
\subsection{Results of $\gamma$-ray Data}
\label{result:gamma}
In our analyses, we find that there are five blazars presenting possible QPOs in their $\gamma$-ray data.
In the following, we give the details on the analyses of the five sources.

We build the maximum likelihood light curves with the time bin of 30 day and calculate the LSP and WWZ power spectra for the 5 objects.
In the constructions of the gamma-ray light curves, only the time bins having TS $>$ 5 are considered.

\begin{figure*}
\centering
	\includegraphics[scale=0.4]{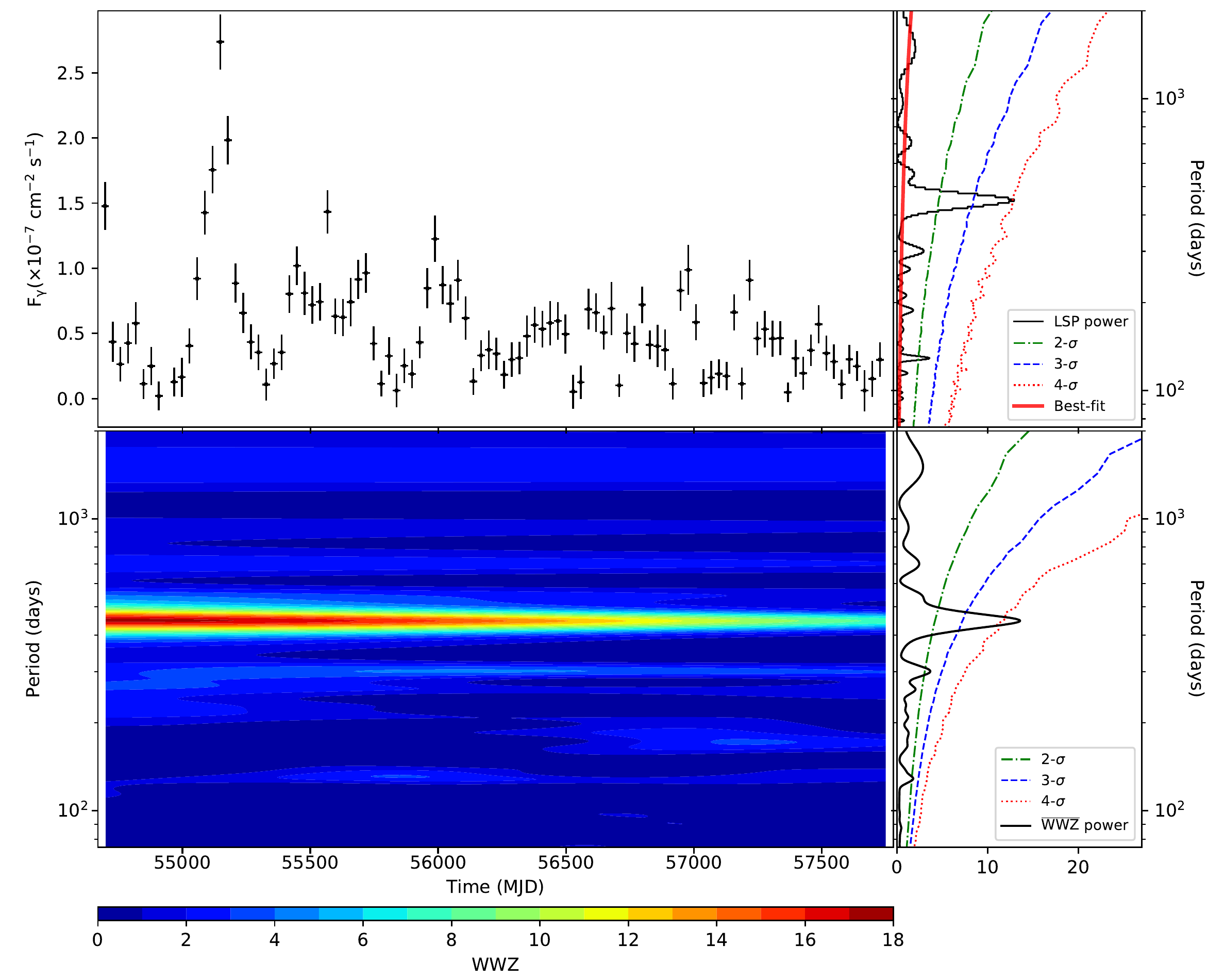}
		\caption{Upper left panel: the 30-day-bin ML $\gamma$-ray light curve of PKS 0601-70 above 100 MeV;
		             Upper right panel: the LSP power spectrum for the light curve shown with black histogram;
		             the red solid line is the best-fit power distribution; and the green dashed-dotted,
		             blue dashed, and red dotted lines represent the 2$\sigma$, 3$\sigma$ and 4$\sigma$
		             simulative confidence levels, respectively.
		             Lower left panel: the 2D colour contour plot of the WWZ power of the light curve.
		             Lower right panel: the time-averaged WWZ power, and the green dashed-dotted, blue dashed, and red dotted lines represent
		             the 2$\sigma$, 3$\sigma$ and 4$\sigma$ confidence levels (for WWZ), respectively.}
	\label{PKS_0601-70}
\end{figure*}

In order to obtain suitable significance of the periodic signal, we model the red noise with a function of power-law plus constant
(their best-fit parameters listed in Table~\ref{result_powerlaw})
by employing a maximum likelihood method \citep{Barret2012}, and then we construct $10^5$ artificial light curves (for each target) with the simulation program provided by \citet{Emmanoulopoulos2013}.
We construct the WWZ spectra using a Morlet mother function \citep{Foster1996} for each artificial light curve.
Then we calculate the probability density distribution of the maxima values of power spectra for LSP and WWZ.  
The significances for the QPOs in the simulated light curves are calculated.
For each target, we also calculate the false-alarm probability (FAP) and the significances corrected by the number of independent frequencies (multiple trials) with the formula,
\begin{equation}
FAP=1-(1-p_{\rm prob})^N,
\label{f2}
\end{equation}
where $N$ is the number of sampled independent frequencies \citep{Zechmeister2009}, which is
calculated with $\frac{f_{\rm max}-f_{\rm min}}{\delta f}$, where $f_{\rm max}$, $f_{\rm min}$ and ${\delta f}$ are respectively its maximum, minimum frequencies, and frequency width.
We find FSRQ PKS 0601-70 presenting quasi-periodic signals with the global significance of $>3\sigma$ in both LSP and WWZ power spectra.
Its power spectra are shown in Fig.~\ref{PKS_0601-70}.
The details on the QPOs for the 5 sources are given in Table~\ref{result_period}.

{\bf PKS 0601-70} is classified as a FSRQ with the redshift of 2.409 \citep{Shaw2012}.
The monthly likelihood light curves and the corresponding LSP and WWZ power spectra are shown in the Fig.~\ref{PKS_0601-70}. 
Obviously its GeV flux variability is violent.
A strong QPO signal with the global significance of $\sim3.9~\sigma$ (~99.99\%) appears near the period of $\sim1.22\pm0.06$ year ($\sim450\pm20$ day).
The uncertainty of the period is evaluated based on the half width at half maximum of the Gaussian fitting.
The significances are shown in Fig.~\ref{PKS_0601-70},
with the green dashed-dotted, blue dashed, and red dotted lines representing the 2$\sigma$, 3$\sigma$ and 4$\sigma$ confidence levels, respectively.

\begin{figure*}
\centering
	\includegraphics[scale=0.4]{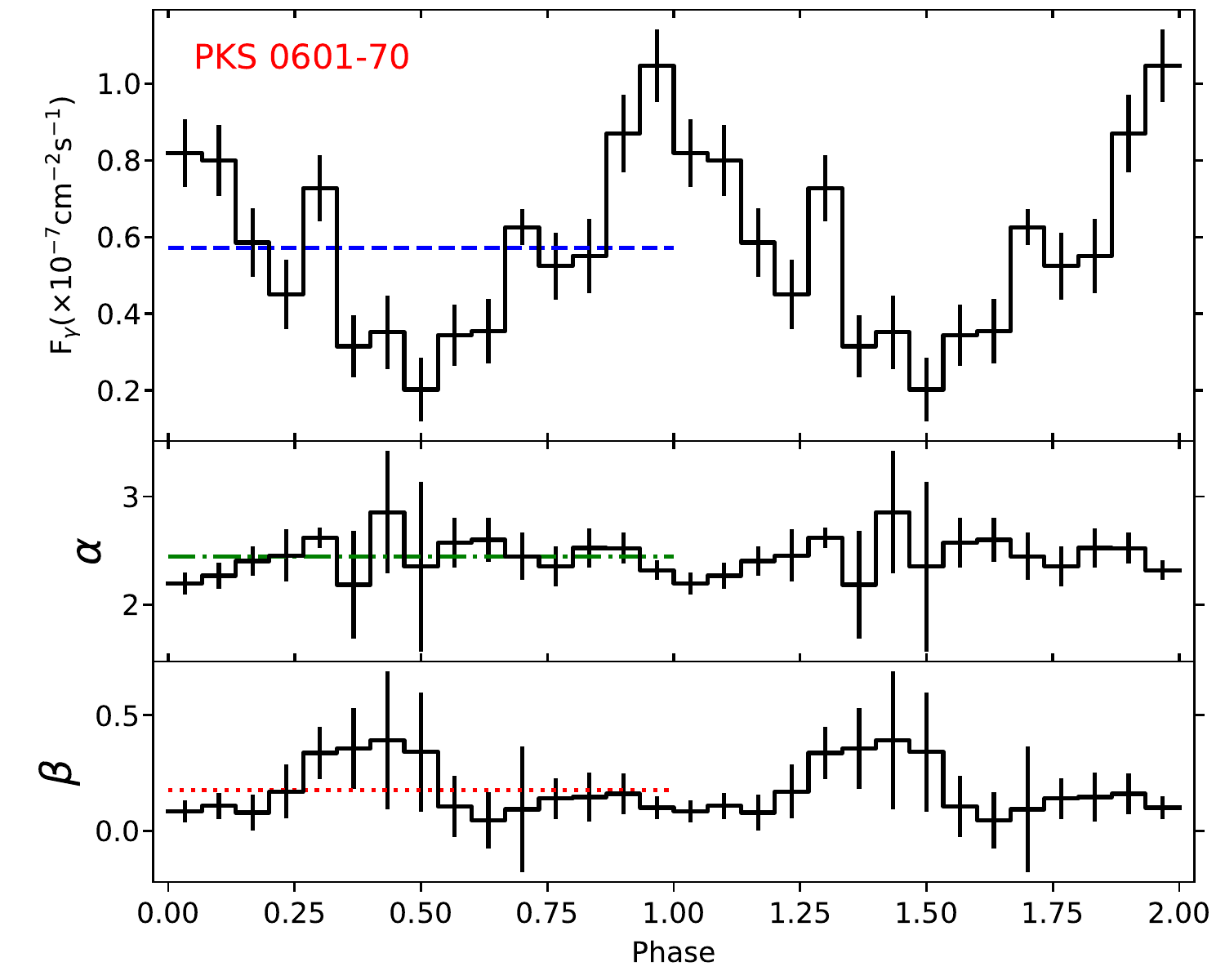}
		\caption{The epoch-folded \emph{Fermi}-LAT observations of PKS 0601-70 above 100 MeV with their period cycles for light curve shape.
		             In the bottom panels $\alpha$ is the photon index in the log-parabola spectrum and $\beta$ is the curvature parameter.
		             And the horizontal lines are their mean values, respectively. For clarity, we show two period cycles.}
	\label{folded_phase}
\end{figure*}

By performing a maximum likelihood fit, we fold the $\gamma$-rays in the ROI centered at its position with the period cycle of 445.6 days to obtain the phase-resolved results of the light curve
and the variability of spectral index (shown in Fig.~\ref{folded_phase}).
We fit the phase-resolved flux and photon indices with constants.
The results (see Table~\ref{result_phase}) indicate that the flux varies with the phase, and the spectral parameters have no significant correlation with the phase (Fig.~\ref{folded_phase}).


For the other four investigated sources (TXS 0518+211, S5 0716+714, B2 2234+28 A, and PKS 0250-225)
we found global statistical significances of periodicity $\sim2\sigma$, too low to single out possible quasi-periodic behaviours.
Information are reported in Tables~\ref{para_source}-\ref{result_phase}.

{\it Note that we analyzed the Fermi-LAT data of the five targets spanning from August 2008 to February 2018. 
The 10-year long light curves cover only several period-cycles.
The caveat again reminds us the complexity of the QPO analysis in AGNs, 
and a similar worry is also proposed in \citet{Covino2019}.}

\subsection{Relation between $\gamma$-ray quasi-periodic frequency and BH mass}
\label{sec:f vs m}
\begin{figure}
\centering
	\includegraphics[scale=0.4]{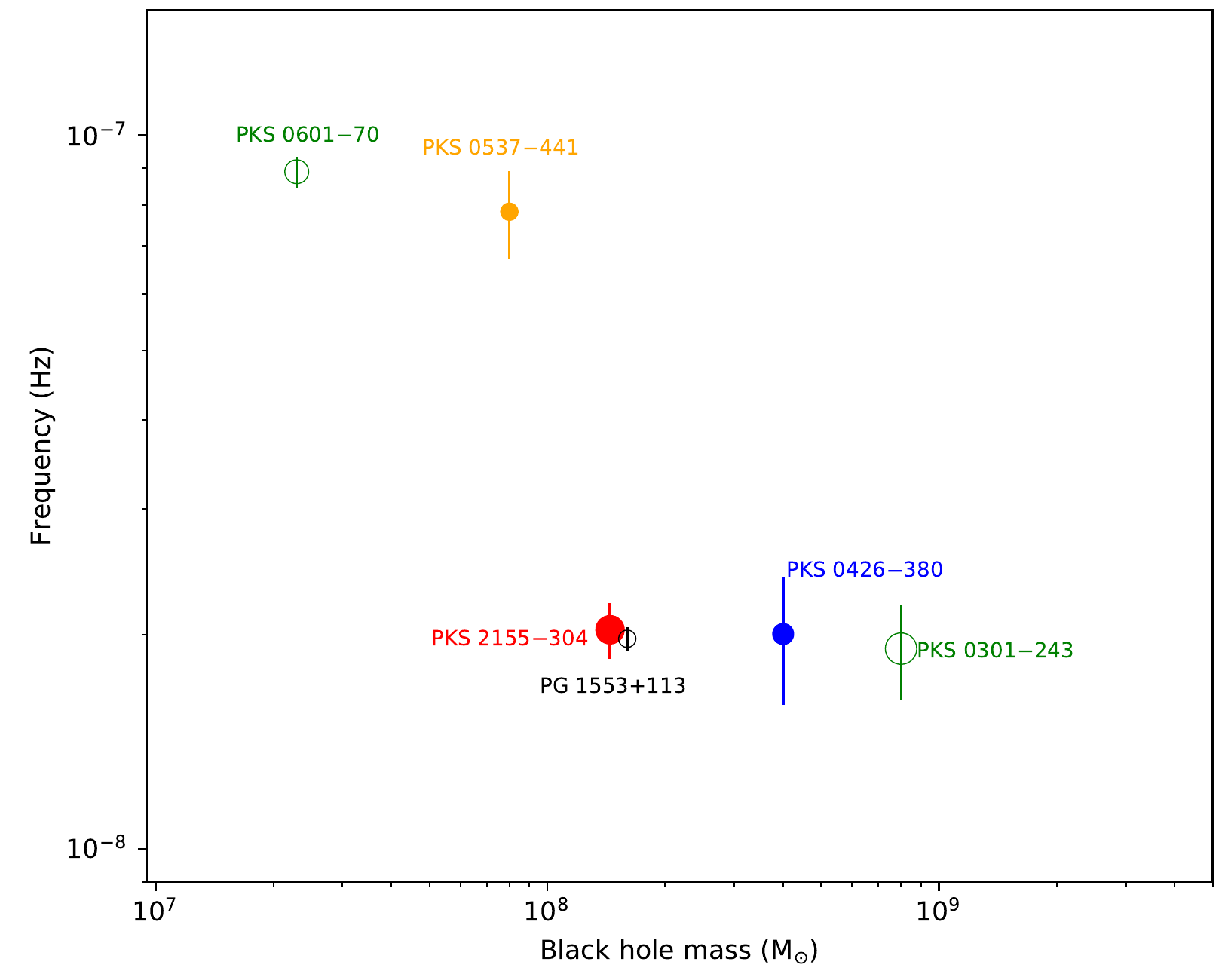}
		\caption{Relationship between the frequency of quasi-periodic signal detected in $\gamma$-ray (corrected by redshift) and the BH mass of AGNs.
		              The results of PKS 2155$-$304, PG 1553+113, PKS 0426$-$380, PKS 0301$-$243 and PKS 0537$-$441
		              are reported in \citet{Sandrinelli2014}, \citet{Zhang2017a}, \citet{1553}, \citet{Zhang2017b}, \citet{Zhang2017c},
		              and \citet{Sandrinelli2016a}. The BH mass of AGNs are listed in Tab.~\ref{result_period}.}
	\label{f_m}
\end{figure}

An inverse correlation between X-ray QPO frequency (corrected by redshift) and BH mass spanning from stellar-mass to supermassive black holes
has been reported in \citet{Kluzniak2002,Remillard2006,Zhou2010,Zhou2015,Pan2016}.
We also investigate the relation between $\gamma$-ray QPO frequency and BH mass of AGNs\footnote{The references for the BH masses are shown in the caption of Fig.~\ref{f_m}.} (see the Fig.~\ref{f_m}).
However no significant correlation is found.

\section{SUMMARY AND DISCUSSION}
\label{sec:summary}

We have carefully analyzed \emph{Fermi}-LAT observations from 2008 to 2018 for 105 AGNs.
 Among them, there are 90 blazars, 10 radio galaxies, and 5 NLSy1s.
Besides the objects reported in previous works, a FSRQ PKS 0601-70 is found having possible quasi-periodic variabilities in its $\gamma$-ray light curves.
Three BL Lacs (TXS 0518+211, S5 0716+714 and B2 2234+28 A), and a FSRQs (PKS 0250-225) present signals with significance of $\sim2\sigma$.

It is interesting that the observed period of PKS 0601-70 is $\sim1.22$ year, 
and the corrected period in the host galaxy frame is $\sim$0.35 year (see Table~\ref{result_period}).
It is also shorter than that reported in the previous works  \citep[e.g.,][]{Sandrinelli2014,Zhang2017a,1553,Zhang2017b,Zhang2017c,Sandrinelli2016a}

\citet{Prokhorov2017} performed a systematic search for QPOs in {\it Fermi}-LAT $\gamma$-ray sources.
They reported seven blazars having possible QPOs.
Among them, the signals in two blazars (PKS 2155-304 and PG 1553+113) have been reported.
The signals in the other 5 blazars are newly discovered.
Among the five blazars, we find no signal in three blazars (PKS 2052-47, BL Lacertae, and PKS 0805-07).
For 4C +01.28, \citet{Prokhorov2017} reported a QPO with the period of 445 days. 
We also find this signal in the data from 2008 August to 2016 December, however this signal disappears in the data from 2008 August to 2018 February.
For S5 0716+714, \citet{Prokhorov2017} reported a QPO with the period of 346 days. Again, we also find this signal in the data from 2008 August to 2016 December, 
however this signal disappears in the data from 2008 August to 2018 February.
We find another signal with the period of $\sim1000\ $days in S5 0716+714, but its global statistical significances is only $2.3\sigma$.

One can see that the result is tentative, and longer/multifrequencies data are needed to confirm these signals.
Besides, we apply a ARIMA model \citep[e.g., ][]{Kelly} to describe the light curves of PKS 0601-70 and its results are shown in Appendix.
The ARIMA results mean that the light curve can also be described by a stochastic process.
It could indicate that it is not necessary to invoke astrophysical mechanism to describe the QPOs \citep[e.g., ][]{Kelly}.
\citet{Covino2019} showed that the power spectral densities computed from the gamma-ray light curves of ten blazars which were claimed having possible QPO 
are all essentially consistent with a red-noise
only. This caveat should be kept in our mind.

Origins of $\gamma$-ray QPOs in AGNs are quite uncertain \citep{Caproni,Covino2017,Sobacchi,Sandrinelli2018,Tavani}.
New results found here could be helpful for us to determine the origins of $\gamma$-ray QPOs.
Their QPOs could be accounted for by (1) quasi-periodic injection of plasma into the jet caused by some kinds of instabilities \citep[e.g.,][]{Tavani};
or (2) periodic change of Doppler factor caused by jet precession \citep[e.g.,][]{Sobacchi}.
The latter origin cannot result in a spectral index oscillation.
In our analysis, the flux and spectral index oscillations are correlated in S5 0716+714.
If it is ture, the correlated flux and spectral index oscillations likely hint that its QPO is intrinsic.
The  MHD-kinetic tearing instabilities in a binary system of SMBHs could produce such intrinsic QPO \citep[e.g.,][]{Tavani}.
Such injection could have impacts both on the power and acceleration rate of high energy electrons,
which then could impact the non-thermal fluxes and spectral indexes.

\citet{yan} proposed a method to test whether the relativistic boost is the cause of gamma-ray QPO in blazar.
Their results support the scenario of the relativistic boost producing the gamma-ray QPO for PG 1553+113.
This method can be applied for these sources to obtain some clues on the origins of the $\gamma$-ray QPOs.

Moreover, two possible reasons accounting for the (non-)correlation between $\gamma$-ray QPO frequency and BH mass (Fig.~\ref{f_m}) are:
(1) the origins of all QPOs are independent of BH mass;  and (2) the QPOs have quite different origins. One can see that origins for $\gamma$-ray QPOs in AGNs may be very complex.
More multiwavelength QPO analysis are needed to resolve the problem of the origins for QPOs in AGNs.

\section*{Acknowledgements}
We thank the referees for helpful and valuable comments.
We acknowledge financial supports from National Key R\&D Program of China under grant No. 2018YFA0404204, 
the National Natural Science Foundation of China (NSFC-U1738124, NSFC-11803081 and NSFC-11661161010), 
and the joint foundation of Department of Science and Technology of Yunnan Province and Yunnan University [2018FY001(-003)].
The work of D. H. Yan is supported by the CAS ``Light of West China'' Program and Youth Innovation Promotion Association.
\bibliography{ApJ}

\begin{thebibliography}{}
\expandafter\ifx\csname natexlab\endcsname\relax\def\natexlab#1{#1}\fi
\bibitem[Abdo et al. (2009)]{Abdo2009}Abdo, A. A., Ackermann, M., Ajello, M., et al. 2009, ApJ, 699, 976
\bibitem[Abdo et al. (2010)]{Abdo2010}Abdo, A. A., Ackermann, M., Ajello, M., et al., 2010, Sci, 328, 725
\bibitem[Atwood et al. (2009)]{Atwood2009}Atwood W. B. et al., 2009, ApJ, 697, 1071
\bibitem[Ackermann et al. (2015)]{1553}Ackermann, M., Ajello, M., Albert, A., et al. 2015, ApJL, 813, L41
\bibitem[Acero et al. (2015)]{Acero2015}Acero, F., Ackermann, M., Ajelloet, M., et al. 2015, ApJS, 218, 23
\bibitem[Bai et al. (1998)]{Bai1998}Bai, J. M., Xie, G. Z., Li, K. H., Zhang, X., \& Liu, W. W. 1998, A\&AS, 132, 83
\bibitem[Bai et al. (1999)]{Bai1999}Bai, J. M., Xie, G. Z., Li, K. H., Zhang, X., \& Liu, W. W. 1999, A\&AS, 136, 455
\bibitem[Barret \& Vaughan (2012)]{Barret2012}Barret, D., \& Vaughan, S. 2012, ApJ, 746, 131
\bibitem[Bhatta et al. (2016)]{Bhatta2016}Bhatta, G., Zola, S., Stawarz, \L., et al. 2016, ApJ, 832, 47
\bibitem[Chai et al. (2012)]{Chai2012}Chai, B., Cao, X., Gu, M., 2012, ApJ, 759, 114
\bibitem[Caproni et al.(2017)]{Caproni}Caproni A., Abraham Z., Motter J. C., \& Monteiro, H. 2017, ApJ, 851, L39
\bibitem[Covino et al. (2017)]{Covino2017}Covino, S., Sandrinelli, A., \& Treves, A. 2017, IAU Symp., 324, 180
\bibitem[Covino et al. (2019)]{Covino2019}Covino, S.; Sandrinelli, A.; Treves, A., 2019, MNRAS, 482, 1279
\bibitem[D'Ammando et al. (2015)]{Da}D'Ammando F., Orienti M., Finke J., et al. 2015, MNRAS, 446, 2456
\bibitem[Emmanoulopoulos et al. (2013)]{Emmanoulopoulos2013}Emmanoulopoulos, D., McHardy, I. M., Papadakis, I. E., 2013, MNRAS, 433, 907
\bibitem[Fan et al. (2011)]{Fan2011}Fan, J.-H., Tao, J., Qian, B.-C., Liu, Y., Yang, J.-H., Pi, F.-P., Xu, W., 2011, RAA, 11, 1311
\bibitem[Finke et al. (2008)]{Finke2008}Finke, J. D., Dermer, C. D., \& B\"ottcher, M. 2008, ApJ, 686, 181
\bibitem[Foster (1996)]{Foster1996}Foster G. 1996, AJ, 112, 1709
\bibitem[Foschini et al. (2015)]{Fo}Foschini L., Berton M., Caccianiga A., et al. 2015, A\&A, 575, A13
\bibitem[Gierli{\'n}ski et al. (2008)]{Gierlinski2008}Gierli{\'n}ski, M., Middleton, M., Ward, M., \& Done, C., 2008, Nature, 455, 369
\bibitem[Horne \& Baliunas (1986)]{HorneBaliunas1986}Horne, J. H. \& Baliunas, S. L., 1986, ApJ, 302, 757
\bibitem[Hong, Xiong \& Bai (2017)]{Hong2017}Hong, Shanwei, Xiong, Dingrong, Bai, Jinming, 2018, AJ, 155, 31
\bibitem[Kelly et al. (2009)]{Kelly}Kelly, B. C., Bechtold, J., \& Siemiginowska, A. 2009, ApJ, 698, 895
\bibitem[Kidger et al. (1992)]{Kidger1992}Kidger, M., Takalo, L., \& Sillanpaa, A. 1992, A\&A, 264, 32
\bibitem[Klu\'zniak \& Abramowicz (2002)]{Kluzniak2002}Klu\'zniak, W., \& Abramowicz, M. A. 2002, arXiv:astro-ph/0203314
\bibitem[Li et al. (2017)]{Li2017}Li Yan-rong, Wang, Jian-min, Zhang, Zhi-xiang, et al., 2017, arXiv:1705.07781
\bibitem[Liao et al. (2015)]{Liao2015}Liao, N.-H., Liang, Y.-F., Weng, S.-S., Gu, M.-F., \& Fan, Y.-Z. 2015, arXiv:1510.05584
\bibitem[Lin et al. (2013)]{Lin2013}Lin, D., Irwin, J. A., Godet, O., et al., 2013, ApJL, 776, L10
\bibitem[Lomb (1976)]{Lomb1976}Lomb, N. R., 1976, Ap\&SS 39, 447
\bibitem[Maraschi et al. (1992)]{Maraschi1992}Maraschi, L., Ghisellini, G., \& Celotti, A. 1992, ApJ, 397, L5
\bibitem[Nilsson et al. (2008)]{Nilsson2008}Nilsson, K., Pursimo, T., Sillanp\"a\"a, A., Takalo, L. O., \& Lindfors, E. 2008, A\&A, 487, L29
\bibitem[Pan et al. (2016)]{Pan2016}Pan, H.-W., Yuan, W.-M., Yao, S., et al. 2016, ApJL, 819, L19
\bibitem[Prokhorov \& Moraghan(2017)]{Prokhorov2017}Prokhorov, D. A. \& Moraghan, A., 2017, MNRAS, 471, 3036
\bibitem[Rani (2010)]{Rani2010}Rani B. et al., 2010, ApJ, 719, L153
\bibitem[Remillard \& McClintock (2006)]{Remillard2006}Remillard, R. A., \& McClintock, J. E. 2006, ARA\&A, 44, 49
\bibitem[Sandrinelli et al. (2014)]{Sandrinelli2014}Sandrinelli, A., Covino, S., \& Treves, A., 2014, ApJL, 793, L1
\bibitem[Sandrinelli et al. (2016a)]{Sandrinelli2016a}Sandrinelli, A., Covino, S., Treves, A., 2016a, ApJ, 820, 20S
\bibitem[Sandrinelli et al. (2016b)]{Sandrinelli2016b}Sandrinelli, A., Covino, S., Dotti, M., Treves, A., 2016b, AJ, 151, 54
\bibitem[Sandrinelli et al. (2017)]{Sandrinelli2017}Sandrinelli, A., Covino, S., \& Treves, A. 2017, A\&A, 600, 132
\bibitem[Sandrinelli et al. (2018)]{Sandrinelli2018}Sandrinelli, A., Covino, S., Treves, A. et al. 2018, A\&A, 615, A118
\bibitem[Scargle (1982)]{Scargle1982}Scargle, J. D., 1982, ApJ, 263, 835
\bibitem[Shaw et al. (2012)]{Shaw2012}Shaw, M. S., Romani, R. W., Cotter, G., et al. 2012, ApJ, 748, 49
\bibitem[Shaw et al. (2013)]{Shaw2013}Shaw, M. S., Romani, R. W., Cotter, G., et al. 2013, ApJ, 764, 135
\bibitem[Sobacchi et al.(2017)]{Sobacchi}Sobacchi E., Sormani,M. C., \& Stamerra A. 2017, MNRAS, 465, 161
\bibitem[Sun et al. (2015)]{Sun}Sun X. N., Zhang J., Lin D. B., et al. 2015, ApJ, 798, 43
\bibitem[Tavecchio et al. (1998)]{Tavecchio1998}Tavecchio, F., Maraschi, L., \& Ghisellini, G. 1998, ApJ, 509, 608
\bibitem[Tavani et al.(2018)]{Tavani}Tavani M., Cavaliere A., Munar-Adrover P., Argan A. 2018, ApJ, 854, 11
\bibitem[Urry \& Padovani (1995)]{Urry1995}Urry, C. M. \& Padovani, P. 1995, PASP, 107, 803 ff.
\bibitem[Urry (2011)]{Urry2011}Urry M., 2011, Journal of Astrophysics and Astronomy, 32, 341
\bibitem[Valtonen et al. (2006)]{Valtonen2006}Valtonen, M. J., Lehto, H. J., Sillanp$\ddot{a}\ddot{a}$, A., et al. 2006, ApJ, 646, 36
\bibitem[Wilman et al. (2005)]{Wilman2005}Wilman, R. J., Edge, A. C., \& Johnstone, R. M., 2005, MNRAS, 359, 755
\bibitem[Yan et al. (2014)]{Yan2014}Yan, D. H., Zeng, H. D., \& Zhang, L. 2014, MNRAS, 439, 2933
\bibitem[Yan et al. (2018)]{yan}Yan, D. H., Zhou, J. N., Zhang, P. F., Zhu, Q. Q., Wang, J. C., 2018,ApJ, 867, 53
\bibitem[Zechmeister \& K\"urster (2009)]{Zechmeister2009}Zechmeister, M. \& K\"urster, M., 2009. A\&A. 496, 577
\bibitem[Zhang et al. (2012)]{Zhang2012}Zhang, B. K., Dai, B. Z., Wang, L. P., Zhao, M., Zhang, L. and Cao, Z., 2012, MNRAS, 421, 3111
\bibitem[Zhang et al. (2014)]{Zhang2014}Zhang B.-K., Zhao X.-Y., Wang C.-X., Dai B.-Z., 2014, Research in Astronomy and Astrophysics, 14, 933
\bibitem[Zhang et al. (2017a)]{Zhang2017a}Zhang, P.-F., Yan, D.-H., Liao, N.-H., Wang, J.-C., 2017a, ApJ, 835, 260
\bibitem[Zhang et al. (2017b)]{Zhang2017b}Zhang, P.-F., Yan, D.-H., Liao, N.-H., Zeng, W., Wang, J.-C., Ciao, L.-J., 2017b, ApJ, 842, 10
\bibitem[Zhang et al. (2017c)]{Zhang2017c}Zhang, P.-F., Yan, D.-H., Zhou, J.-N., Fan, Y.-Z., Wang, J.-C., Zhang, L., 2017c, ApJ, 845, 82
\bibitem[Zhang et al. (2017d)]{Zhang2017d}Zhang, P., Zhang, P. F., Yan, J. Z., et al. 2017d, ApJ, 849, 9
\bibitem[Zhang et al. (2018)]{Zhang2018}Zhang, P. F., Zhang, P., Liao, N. H., et al., 2018, ApJ, 853, 193
\bibitem[Zhang et al. (2017)]{zhangjin}Zhang, J., Zhang, H. M., Zhu, Y. K., Yi, T. F., Yao, S., Lu, R. J., Liang, E. W., 2017, ApJ, 849, 42
\bibitem[Zhou et al. (2010)]{Zhou2010}Zhou, X.-L., Zhang, S.-N., Wang, D.-X., \& Zhu, L. 2010, ApJ, 710, 16
\bibitem[Zhou et al. (2015)]{Zhou2015}Zhou, X.-L., Yuan, W.-M., Pan, H.-W., \& Liu, Z. 2015, ApJL, 798, L5
\end{thebibliography}
\appendix
\section{Autocorrelation analysis and $\gamma$-ray QPO in blazar}
In addition to the methods based on Fourier analysis (e.g., LSP and WWZ), 
an autocorrelation analysis also has been used in light curve analysis to search for quasi-periodic behavior
or variations with a characteristic timescale \citep{Kelly}.

\begin{figure}
\centering
	\includegraphics[scale=0.4]{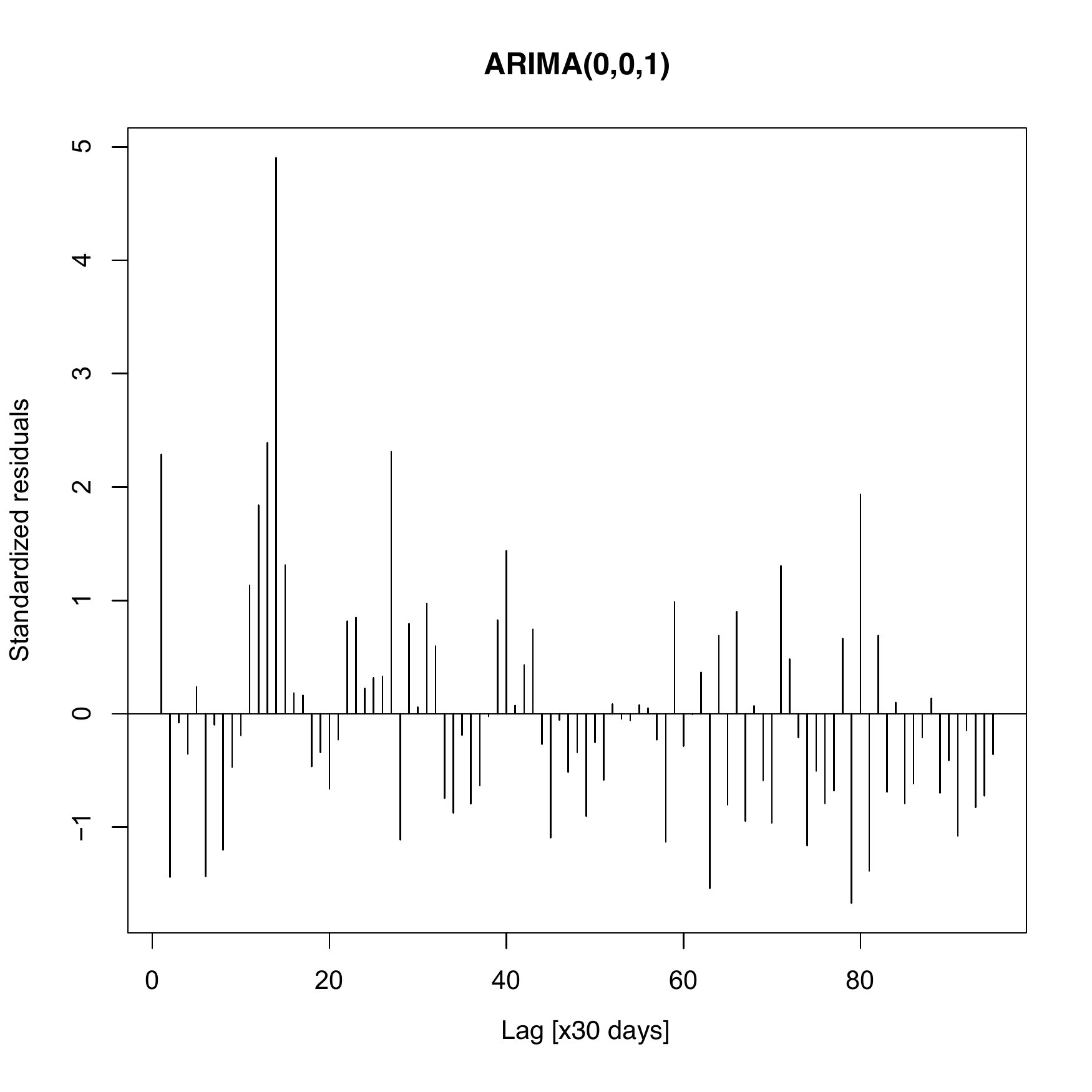}
	\includegraphics[scale=0.4]{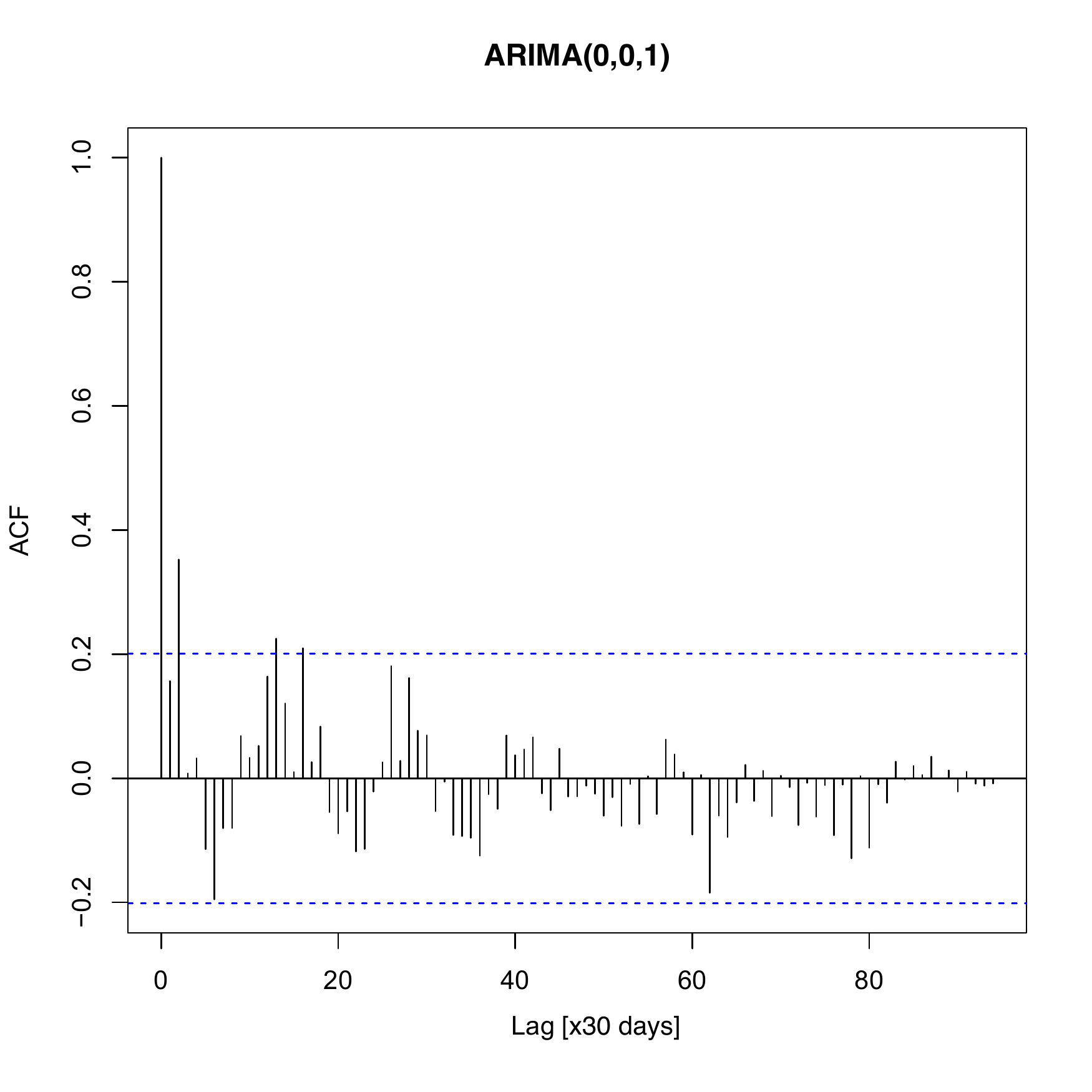}
		\caption{Results of ARIMA model fitting for the monthly likelihood gamma-ray light curve of PKS 0601-70.
		              Left panel: standard residuals of the ARIMA(0,0,1) model fitting.
		              Right panle: residuals ACF of the ARIMA(0,0,1) model fitting, where the dashed horizontal lines represent the 95\% confidence level.}
	\label{PKS_0601-70_arima}
\end{figure}

We use the ARIMA(0,0,1) model to fit the light curves of the five blazars.
The light curves of the five blazars can be fitted well by the ARIMA(0,0,1) model, which is a stochastic process model.
For examples, the results of PKS 0601-70 are shown in Fig.~\ref{PKS_0601-70_arima}.
This means that the light curves can be described by a stochastic process. 
It could indicate that it is not
necessary to invoke astrophysical mechanism to
describe the QPOs \citep[e.g.,][]{Kelly}.
\clearpage

\begin{sidewaystable}[h]
\begin{center}
\caption{Results for the five AGNs information.}
\begin{tabular}{c|rr|rr|c|c|c|cc|r|r|c}
\hline\hline
\multirow{2}{*}{Source name$^{(1)}$}& \multicolumn{4}{c|}{Centre of ROI$^{(2)}$} &Model$^{(3)}$ & Redshift$^{(4)}$  & \multirow{2}{*}{3FGL$^{(5)}$} & \multicolumn{2}{c|}{\multirow{2}{*}{Index$^{(6)}$}}& Flux$^{(7)}$~~~~ &  \multirow{2}{*}{TS$^{(8)}$~~} & Energy band$^{(9)}$\\
\cline{2-7}\cline{11-11}\cline{13-13}
                  &R.A.~~& decl.~~&$l$~~~~~& $b$~~~~~& (in 3FGL) & ($z$)  & & & & ($\times10^{-8})~~~~$ & &(GeV) \\
\hline
PKS 0601$-$70	  & 90.313 & $-$70.609 & 281.042 & $-$29.624 & LogParabola & 2.409 &3FGL J0601.2$-$7036 & $2.354\pm0.034$ & $0.405\pm0.033$ & $5.72\pm0.25$ & 2715.1 & 0.1$-$500\\
\hline
PKS 0250$-$225    & 43.201 & $-$22.311 & 209.723 & $-$62.091 & LogParabola & 1.419 &3FGL J0252.8$-$2218 & $2.217\pm0.022$ & $0.085\pm0.012$ & $8.44\pm0.19$ & 9677.7 & 0.1$-$500\\
TXS 0518$+$211	  & 80.443 & 21.222 & 183.599 & $-$8.702 & PowerLaw & 0.108 &3FGL J0521.7$+$2113 & \multicolumn{2}{c|}{$1.922\pm0.018^{\dag}$} & $9.90\pm0.42$ & 13851.7 & 0.1$-$500\\
S5 0716$+$714     & 110.490  & 71.345 & 143.980 & 28.023 & LogParabola & 0.300  &3FGL J0721.9$+$7120 & $1.999\pm0.008$ & $0.031\pm0.003$ & $22.85\pm0.20$ & 95738.6 & 0.1$-$500\\
B2 2234+28 A   & 339.077 & 28.489 & 90.108 & $-$25.635 & LogParabola & 0.790 &3FGL J2236.3$+$2829 & $2.115\pm0.029$ & $0.108\pm0.015$ & $6.06\pm0.20$ & 5870.0 & 0.1$-$500\\
\hline\hline
\end{tabular}
\label{para_source}
\end{center}
{\bf Notes.}
(1)Identifier for targets. (2)Coordinates of ROI centre in data analysis derived from 3FGL in J2000 and their Galactic coordinates.
(3)The energy spectral shape in model file. (4)Redshift. (5)The $\gamma$-ray source designations in 3FGL.
(6)The best-fitting of photon spectral index derived from maximum likelihood analysis for 0.1$-$500 GeV. $^{\dag}$Only for TXS 0518$+$211 having energy spectrum type of PowerLaw.
(7)The integrated photon flux above 0.1 GeV and its 1$\sigma$ uncertainty with units of photons/cm$^2$/s. (8)The likelihood Test Statistic value. (9) Energy band in events selection.
\end{sidewaystable}
\begin{table*}
\begin{center}
\caption{Best-fitting results of power spectra for five AGNs.}
\begin{tabular}{c|c|c|c|c|c|c}
\hline\hline
\multirow{2}{*}{Source name}& \multirow{2}{*}{Class}&\multicolumn{3}{c|}{Parameters of power spectra$^{(1)}$} &\multirow{2}{*}{$\Delta\chi^{2~(2)}$} & Simulation$^{(3)}$ \\
\cline{3-5}\cline{7-7}
               & & $\log(A)$ & $\alpha$ & $\log(C)$ & &($times$) \\
\hline
PKS 0601$-$70  & FSRQ&  $-1.81\pm0.57$ & $0.61\pm0.22$ & $-2.95\pm1.60$ & 0.3 & $\sim10^5$ \\
\hline
PKS 0250$-$225 & FSRQ& $-1.68\pm0.49$ & $0.62\pm0.21$ & $-2.99\pm1.88$ & 0.3 & $\sim10^5$ \\
TXS 0518$+$211 & BL Lac& $-3.98\pm1.52$ & $1.33\pm0.53$ & $-0.72\pm0.18$ & 0.3 & $\sim10^5$ \\
S5 0716$+$714  & BL Lac&  $-2.47\pm0.64$ & $0.90\pm0.25$ & $-2.69\pm2.32$ & 0.2 & $\sim10^5$ \\
B2 2234+28 A & BL Lac&  $-2.24\pm0.50$ & $0.82\pm0.21$ & $-3.01\pm2.18$ & 0.4 & $\sim10^5$ \\
\hline\hline
\end{tabular}
\label{result_powerlaw}
\end{center}
{\bf Notes.}
(1)The best-fitting results of power spectra of gamma-ray light curves. (2)Reduced $\chi^2$/d.o.f with d.o.f $\sim$ 130. (3)The times for simulation.
\end{table*}
\begin{table*}
\begin{center}
\caption{Results for $\gamma$-ray quasi-periodic signals of five AGNs.}
\begin{tabular}{c|c|c|c|c}
\hline\hline
\multirow{2}{*}{Source name}& Period cycle$^{(1)}$ & Corrected period$^{(2)}$ & \multirow{2}{*}{Significances}$^{(3)}$& Mass$^{(4)}_{\rm[ref.^{(5)}]}$ \\
\cline{2-3}\cline{5-5}
               & (year) & (year) & & $Log(M_{BH}/M_{\odot})$ \\
\hline
PKS 0601$-$70  & $1.22\pm0.06$ & $0.35\pm0.02$  &3.9$\sigma$ & 7.4$_{\rm[S12]}$\\
\hline
PKS 0250$-$225 & $1.25\pm0.05$ & $0.52\pm0.02$  &2.6$\sigma$  & 9.4$_{\rm[S12]}$\\
TXS 0518$+$211 & $3.1\pm0.4$ & $2.8\pm0.4$ & 2.1$\sigma$  &8.8$_{\rm[S13]}$ \\
S5 0716$+$714  & $2.6\pm0.4$  & $2.0\pm0.3$  &2.3$\sigma$& 7.9$_{\rm[F11,Z12,C12]}$ \\
B2 2234+28 A & $1.31\pm0.05$ &$0.73\pm0.03$ &2.6$\sigma$ & 8.4$_{\rm[S12]}$\\
\hline\hline
\end{tabular}
\label{result_period}
\end{center}
{\bf Notes.}
(1)The observed period cycle (with error) derived from gamma-ray light curves. (2)The intrinsic orbital time (the observed period corrected by redshift).
(3)The multiple trail significances calculated by formulae (\ref{f2}).
(4)BH mass for each target.
(5)References for BH mass. F11: \citet{Fan2011}; Z12: \citet{Zhang2012}; C12: \citet{Chai2012}; S12: \citet{Shaw2012}; S13: \citet{Shaw2013}; W05: \citet{Wilman2005}.
\end{table*}
\begin{table*}
\begin{center}
\caption{Phase-resolved fitting results for the five AGNs.}
\begin{tabular}{c|cc|cc|cc|cc|cc|cc}
\hline\hline
\multirow{2}{*}{Source name}& \multicolumn{6}{c|}{Mean values$^{(1)}$} & \multicolumn{6}{c}{$\chi^2_{\rm min}$/d.o.f; (d.o.f = 14)$^{(2)}$} \\
\cline{2-13}
                                               & \multicolumn{2}{c|}{Flux ($\times10^{-8}$)} &\multicolumn{2}{c|}{$\alpha$} & \multicolumn{2}{c|}{$\beta$} &\multicolumn{2}{c|}{Flux} &\multicolumn{2}{c|}{$\alpha$} & \multicolumn{2}{c}{$\beta$} \\
\hline
PKS 0601$-$70  & \multicolumn{2}{c|}{5.714} & \multicolumn{2}{c|}{2.445} & \multicolumn{2}{c|}{0.177} & \multicolumn{2}{c|}{104.170} & \multicolumn{2}{c|}{16.041} & \multicolumn{2}{c}{15.089} \\
\hline
PKS 0250$-$225 & \multicolumn{2}{c|}{8.254} & \multicolumn{2}{c|}{2.241} & \multicolumn{2}{c|}{0.111} & \multicolumn{2}{c|}{359.516} & \multicolumn{2}{c|}{14.672} & \multicolumn{2}{c}{23.946} \\
\cline{4-7}\cline{10-13}
TXS 0518$+$211 & \multicolumn{2}{c|}{$9.902$} & \multicolumn{4}{c|}{1.930} & \multicolumn{2}{c|}{249.414} & \multicolumn{4}{c}{21.878} \\
\cline{4-7}\cline{10-13}
S5 0716$+$714  & 23.144& 23.449& 2.012 & 2.017 & 0.033 & 0.035& 572.510 &1061.967 & 50.877 & 86.406 & 9.165 &20.839 \\
B2 2234$+$28 A & \multicolumn{2}{c|}{6.308} & \multicolumn{2}{c|}{2.138} & \multicolumn{2}{c|}{0.110} & \multicolumn{2}{c|}{105.484} & \multicolumn{2}{c|}{11.524} & \multicolumn{2}{c}{8.019} \\
\hline\hline
\end{tabular}
\label{result_phase}
\end{center}
{\bf Notes.}
(1)Mean values of flux and indices for the phase-resolved maximum likelihood results (Flux with units of photons/cm$^2$/s).
(2)Reduced $\chi^2$/d.o.f (d.o.f = 14) for the fitting with a constant.
\end{table*}
\end{document}